\documentclass[11.5pt]{article}
\usepackage{graphicx}
\usepackage{amssymb}
\usepackage{amstext}
\usepackage{amsbsy}
\usepackage[normalem]{ulem}
\usepackage{amsmath}
\usepackage{graphicx}
\usepackage{floatflt}
\usepackage{float}
\usepackage{wrapfig}
\numberwithin{equation}{section}

\begin{document}

\huge
\begin{center}
The effect of surface tension on steadily translating bubbles in an unbounded Hele-Shaw cell
\end{center}

\vskip 0.125truein
\large
\begin{center}
Christopher C Green$^{1}$, Christopher J Lustri$^{2}$ and Scott W McCue$^{1}$	
\end{center}

\normalsize
\begin{center}
$^{1}$School of Mathematical Sciences, Queensland University of Technology, Brisbane QLD 4000, Australia. \\
$^{2}$Department of Mathematics, Macquarie University, Sydney NSW 2109, Australia.
\end{center}

\begin{abstract}
New numerical solutions to the so-called selection problem for one and two steadily translating bubbles in an unbounded Hele-Shaw cell are presented. Our approach relies on conformal mapping which, for the two-bubble problem, involves the Schottky-Klein prime function associated with an annulus. We show that a countably infinite number of solutions exist for each fixed value of dimensionless surface tension, with the bubble shapes becoming more exotic as the solution branch number increases. Our numerical results suggest that a single solution is selected in the limit that surface tension vanishes, with the scaling between the bubble velocity and surface tension being different to the well-studied problems for a bubble or a finger propagating in a channel geometry.


\end{abstract}

\section{Introduction}

\vskip 0.1truein
\noindent
A Hele-Shaw cell is a classical experimental apparatus which sandwiches a thin layer of viscous fluid between two parallel glass plates in such a way that the fluid dynamics can be modelled as completely two dimensional.  From a modelling perspective, if the fluid is assumed to be incompressible and the flow irrotational, any mathematical solutions will be necessarily harmonic.  For this reason, flow in a Hele-Shaw cell can be used to produce good visualisations of the streamline patterns in potential flow fields, such as the flow around a circular cylinder or around a body with sharp corners, since the fluid is able to negotiate corners without separation (Van Dyke \cite{vanDyke}).

A particularly interesting set-up involves two immiscible fluids of differing viscosity where, if the less viscous fluid is displacing the more viscous fluid, the interface between the two fluids is unstable and can evolve to form one or more fingers \cite{ST}.  Mathematically, Hele-Shaw models with fluid interfaces give rise to many interesting free boundary problems which have been well-studied over the past several decades, each of which come under the general umbrella of Laplacian growth processes.  Many exact and numerical solutions have been found to these problems, in both steady and unsteady cases, which is particularly remarkable given the highly non-linear nature of these problems (see, for example, Gustafsson \& Vasil'ev \cite{BjornBook}).

We are interested in steadily propagating bubbles in a Hele-shaw cell, a topic which dates back to Taylor \& Saffman~\cite{TS}.  We will focus on the theoretical situation where the Hele-Shaw cell is unbounded, therefore completely removing the effect of any side walls on any of the bubbles and flow field variables.  The standard zero surface tension model for Hele-Shaw flow predicts that any number of bubbles can translate steadily with a continuum of possible bubble speeds $U$ for a given background fluid speed $V$ in the same direction (where $U>V)$.  In the case of a single bubble in an unbounded Hele-Shaw cell, it is relatively simple to show that the bubbles are elliptic in shape with aspect ratio $(U-V)/V$.  For multiple bubbles, exact solutions have been constructed by Crowdy~\cite{Crowdy2009} by exploiting the function theory of the Schottky-Klein prime function \cite{Crowdy2008,computSK}, an approach which is crucial for solving problems like this with arbitrary finite connectivity.  As with the single bubble problem, solutions exist for a continuum of bubble speeds $U$.

There is a similar history for zero surface tension problems in a channel geometry (where the flow is confined to a channel whose parallel walls are perpendicular to the Hele-Shaw plates). Here, in the case of a single bubble in a Hele-Shaw channel which is reflectionally symmetric about the channel centreline, Taylor \& Saffman \cite{TS} appealed to hodograph methods in simply connected domains to exactly determine the shape of the propagating bubble. Tanveer \cite{Tanveer87} later found a wider class of possible bubble shapes in this channel geometry by relaxing the centerline symmetry constraint thereby introducing a doubly connected flow domain and writing the solutions in terms of elliptic functions.  More recently, Green \& Vasconcelos \cite{GreenVas14} found all possible bubble shapes for a steadily translating assembly of finitely many channel bubbles, without any symmetry assumptions, in which the solutions of Taylor \& Saffman \cite{TS} and Tanveer \cite{Tanveer87} are all special cases (the results in \cite{GreenVas14} were further generalised in \cite{Vasconcelos2015} to hold for any number of bubbles and fingers).  While we are concerned here with steady motion of bubbles, recent time-dependent studies for a single bubble in the zero surface tension context are also worth mentioning \cite{Alimov2016,Khalid2015,VMW2014}.

In this paper, we shall be analysing the effect of adding surface tension to the bubble boundaries, thereby rendering the physical model of the flow more precise (we shall denote the dimensionless surface tension by $B$). The addition of surface tension adds a layer of mathematical richness to the problem in that it allows for a countably infinite family of solutions, each with a distinct speed $U$, instead of a continuum of possible bubble speeds.  We shall explore this solution behaviour for the case of one or two bubbles, covering the simply connected and doubly connected cases.  We are interested in using a combination of complex variable techniques, conforming mappings and numerical methods to determine bubble shapes for each of the families of solutions and to predict which of the continuum of solutions is selected in the limit that $B\rightarrow 0$. Our work is complementary to that of Combescot \& Dombre \cite{Combescot} and Tanveer \cite{Tanveer86,Tanveer87,Tanveer89} who considered the selection problem for a single bubble in a channel geometry using both numerical method and exponential asymptotics. It is worth noting that these authors derive the scaling $U/V\sim 2- k B^{2/3}$, as $B\rightarrow 0$, where $k$ is a constant that depends on the solution branch number.  Further, as mentioned by Tanveer~\cite{Tanveer89}, the limits of vanishing bubble area and vanishing surface tension do not commute in the channel geometry, so this scaling does not necessarily apply in the unbounded case (that we consider here).

Our framework for bubble selection is analogous to the well-studied problem of a steadily travelling finger (a Saffman-Taylor finger) in a Hele-Shaw channel, which also gives rise to exact solutions for the zero surface tension case with a finger speed $U$ left as a free parameter (Saffman \& Taylor \cite{ST}). Numerical approaches \cite{McLeanSaffman,VDB1983,GardinerEtAlb} and asymptotic analysis \cite{CombescotEtAl,Chapman99,Tanveer872,Hong,Shraiman} have been applied to the selection problem with surface tension which also leads to a countably infinite number of finger solutions for non-zero surface tension and the selection of $U/V=2$ in the limit that surface tension vanishes. Other selection mechanisms such as anisotropic surface tension \cite{Dorsey}, kinetic undercooling \cite{ChapmanKing,GardinerEtAl2,Dallaston}, or even selection without additional physics \cite{Mineev}, have been explored in some detail and also predict that the $U/V=2$ solution is the physically relevant case, as reported in the original experiments by experiments Saffman \& Taylor \cite{ST}.  Note that all these cases involve simply connected geometries.

The main purpose of this work is to present new numerical solutions to the selection problems for a single steadily translating bubble and for a pair of up-down symmetric steadily translating bubbles in an unbounded Hele-Shaw cell.  The two-bubble problem involves a dimensionless separation distance between the bubbles, so that in the limit of infinitely large separation distance, the two-bubble problem reduces to the simpler single bubble case.  For both the one-bubble and two-bubble problems, we compute numerical solutions for a number of solution branches, demonstrating that the shapes of the bubbles become increasingly exotic as the branch number increases.  Further, our numerical results suggest that $U/V\sim 2-kB^{2}$ as $B\rightarrow 0$, confirming the selection of $U/V=2$ but providing a different surface tension scaling to the channel case.

\section{One bubble}
\noindent
We begin with the treatment of one steadily translating bubble in an unbounded Hele-Shaw cell. The formulation of this problem will pave the way to a generalisation to include two bubbles which we will consider in the proceeding section.

\subsection{Problem formulation}
\noindent
Consider a single bubble in steady motion in an unbounded Hele-Shaw cell. Let $D$ be the unbounded planar region, in the $z=x+\mathrm{i}y$-plane, containing incompressible fluid exterior to a single bubble of finite area. We assume the bubble is reflectionally symmetric about the real axis $y=0$ (on physical grounds, it makes no sense to consider asymmetric bubbles in an unbounded Hele-Shaw cell). Denote the bubble by $D_0$ and its boundary by $\partial D_0$. We will assume that the bubble is moving with constant speed $U$ in the $x$-direction, and that the fluid velocity far away from the bubble is in the $x$-direction with constant speed $V$. The velocity field $\mathbf{u}$ is derived from a velocity potential $\phi$ so that $\mathbf{u}=\nabla \phi$. Since $\phi$ is proportional to the fluid pressure in the Hele-Shaw system, it must be the case that $\phi$ is single-valued everywhere in $D$. Also let $\psi$ denote the streamfunction for this flow. The governing equation to be solved, owing to incompressibility $\nabla \cdot \mathbf{u}=0$, is Laplace's equation for $\phi$, which must be solved in the domain $D$. The following must therefore hold:
\begin{align}
\nabla^2 \phi &=0, \quad z \in D; \\
\phi &= -\frac{b^2}{12 \mu} \sigma \kappa + \phi_0, \quad z \in \partial D_0; \\
v_n &= \frac{\partial \phi}{\partial n}, \quad z \in \partial D_0; \\
\phi &\sim V x, \quad |z|\rightarrow \infty.
\end{align}
Here, $b$ is the gap width between plates, $\mu$ is the fluid viscosity, $\sigma$ is the surface tension of the bubble, $\kappa$ is the curvature of the bubble boundary, and $\phi_0$ is a real constant. Moving to a frame of reference co-travelling with the bubble at speed $U$, the above will change to the following (where the tilde notation denotes quantities in this co-travelling reference frame):
\begin{align}
\nabla^2 \tilde \phi &=0, \quad z \in D; \\
\tilde \phi &= -\frac{b^2}{12 \mu} \sigma \kappa + \phi_0 - U x, \quad z \in \partial D_0; \\
\tilde \psi &= 0, \quad z \in \partial D_0; \\
\tilde \phi &\sim (V-U) x, \quad |z|\rightarrow \infty.
\end{align}
The third of these equations comes about because, in this co-travelling frame, the bubble boundary is a streamline, and without loss of generality, we may fix the value of the streamfunction to be zero here. Let us scale lengths with respect to $L$ and speeds with respect to $V$ (say). Then we have the following dimensionless, hatted quantities:
\begin{equation}
\hat x = \frac{x}{L}, \quad \hat U=\frac{U}{V}, \quad \hat \phi =\frac{\tilde \phi}{VL}, \quad \hat \psi = \frac{\tilde \psi}{VL}, \quad \hat \kappa = L \kappa.
\end{equation}
We then have the following non-dimensionalised problem:
\begin{align}
\nabla^2 \hat \phi &=0, \quad \hat z \in D; \\
\hat \phi &= -\frac{b^2}{12 \mu V L^2} \sigma \hat \kappa + \hat \phi_0 - \hat U \hat x, \quad \hat z \in \partial D_0;\\
\hat \psi &= 0, \quad \hat z \in \partial D_0;\\
\hat \phi &\sim (1-\hat U ) \hat x, \quad |\hat z|\rightarrow \infty.
\end{align}
We note from the above that we may define a non-dimensional surface tension parameter through
\begin{equation}
B=\frac{b^2 \sigma}{12 \mu V L^2}.
\end{equation}
We assume that for a convex bubble, the curvature of the bubble boundary $\kappa$ will always be strictly positive.

\vskip 0.1truein
\noindent
We may choose $L^2$ to be the bubble area $\tilde A$ divided by $\pi$. Then the dimensionless area
\begin{equation}
\hat A = \frac{\tilde A}{L^2} = \pi
\end{equation}
(say) so that
\begin{equation}
L=\sqrt{\frac{\tilde A}{\pi}} \quad \mathrm{and} ~~ B=\frac{\pi b^2 \sigma}{12 \mu V \tilde A}.
\label{scalings}
\end{equation}
We will now drop the hatted notation and assume we are working with dimensionless variables henceforth. We thus have the following problem to solve, for the velocity potential $\phi$ and the streamfunction $\psi$, pertaining to a single bubble in an unbounded Hele-Shaw cell in a frame of reference co-travelling with the bubble:
\begin{align}
\nabla^2 \phi &=0, \quad z \in D; \label{goveq1} \\
\phi + U x &= B \kappa + \phi_0, \quad z \in \partial D_0; \label{goveq2} \\
\psi &= 0, \quad z \in \partial D_0; \label{goveq3} \\
\phi &\sim (1-U) x, \quad |z|\rightarrow \infty. \label{goveq4}
\end{align}
In the above, $\phi_0$ is a real constant. It makes sense, thus, to solve for a complex potential function $w(z)=\phi+\mathrm{i}\psi$ for this flow whose real and imaginary parts are respectively the velocity potential $\phi$ and the streamfunction $\psi$. Function $w$ must be a single-valued analytic function everywhere in $D$. A natural way of doing this is to pull-back to a simpler parametric $\zeta$-plane and proceed to find the composition $W(\zeta)=w(z(\zeta))$, where $z(\zeta)$ is a conformal map to the fluid region $D$ exterior to the bubble, and must also be determined. Label the pre-image region of $D$ in the $\zeta$-plane by $D_\zeta$ which may be taken to be the unit $\zeta$-disc without loss of generality. Then let $C_0$ label the unit $\zeta$-circle, $|\zeta|=1$. The Riemann mapping theorem guarantees the existence of a conformal mapping $z(\zeta)$ between the two simply connected regions $D_\zeta$ and $D$. We will now re-write the problem (\ref{goveq1})-(\ref{goveq4}) in terms of the complex functions $z(\zeta)$ and $W(\zeta)$, or equivalently, in terms of the complex variable $\zeta$.

\begin{figure}
\begin{center}
\includegraphics[scale=0.3]{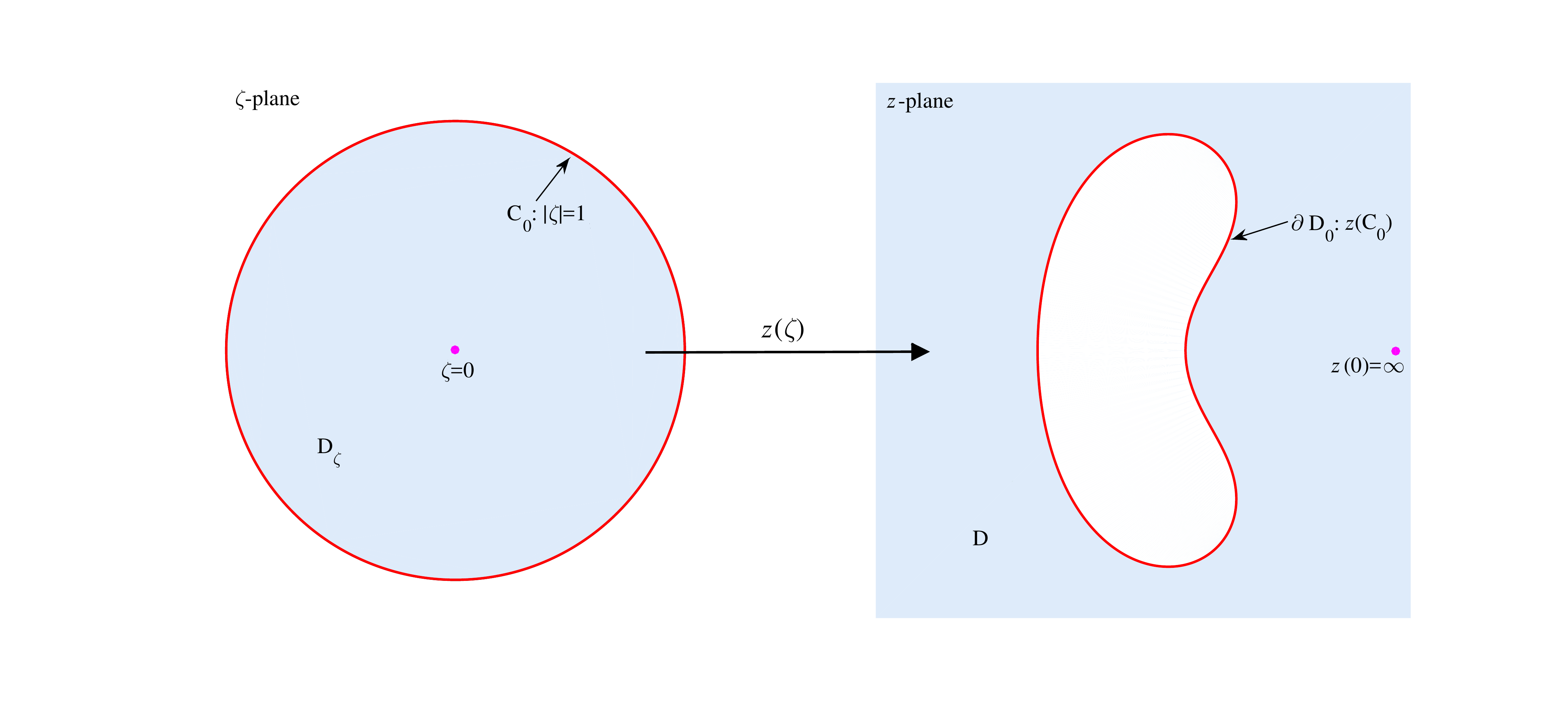}
\vskip -0.3truein
\caption{Schematic of the pre-image parametric $\zeta$-plane being the interior of the unit circle $C_0$ (left), and the target image domain in the $z$-plane exterior to the bubble (right).}
\label{schematic1}
\end{center}
\end{figure}

\vskip 0.1truein
\noindent
In the frame of reference co-travelling with the bubble, the complex potential $w(z)$ is related to the complex potential $\tilde w(z)$ in the laboratory frame via $w(z)=\tilde w(z)-Uz$. It follows that $w$ is a single-valued analytic function everywhere in $D$ except for a simple pole singularity at infinity. Equivalently, $W(\zeta)$ is a single-valued analytic function everywhere in $D_\zeta$ except for a simple pole singularity at the point $\zeta=\beta \in D_\zeta$ (say) mapping to infinity. Correspondingly, the conformal map $z(\zeta)$ must have a simple pole also at $\zeta=\beta$ but is otherwise analytic and single-valued. This means that locally, we must have
\begin{equation}
z(\zeta) \sim \frac{a}{\zeta-\beta}, \quad \zeta \rightarrow \beta,
\label{zbehav}
\end{equation}
for some constant $a$, which can be taken to be real; this leaves two remaining real degrees of freedom associated with the Riemann mapping theorem which allows us to choose $\beta=0$ (say). We will construct $z(\zeta)$ such that $C_0$ maps to the bubble boundary $\partial D_0$. Figure \ref{schematic1} shows a schematic.

\vskip 0.1truein
\noindent
It is a trivial observation that if $W(\zeta)=\Phi + \mathrm{i} \Psi$ is a single-valued analytic function of $\zeta$, then both its real and imaginary parts $\Phi$ and $\Psi$ are necessarily harmonic, as required.
(\ref{goveq2}) becomes
\begin{equation}
\mathrm{Re}[W(\zeta)+Uz(\zeta)]=B\kappa+\mathrm{constant}, \quad \zeta \in C_0,
\end{equation}
where an expression for the signed curvature can be given in terms of the conformal map $z(\zeta)$ through the formula
\begin{equation}
\kappa = \pm \left(\frac{1+\mathrm{Re}[\zeta z''(\zeta)/z'(\zeta)]}{|z'(\zeta)|}\right).
\label{curvature}
\end{equation}
We will adopt the following sign convention. We wish to measure the curvature of the bubble boundary from a point lying within the interior of the bubble. Since our conformal map reverses the orientation of a tangent vector in the $\zeta$-plane, we thus require a \textit{minus} sign in (\ref{curvature}), so that $\kappa>0$ when the tangent vector, directed out of the domain, rotates anti-clockwise. (\ref{goveq3}) is simply
\begin{equation}
\mathrm{Im}[W(\zeta)]=0, \quad \zeta \in C_0.
\end{equation}
(\ref{goveq4}) transforms to
\begin{equation}
W(\zeta) \sim (1-U)z(\zeta), \quad \zeta \rightarrow \beta,
\end{equation}
which, in light of (\ref{zbehav}) and the choice $\beta=0$, is
\begin{equation}
W(\zeta) \sim \frac{(1-U)a}{\zeta}, \quad \zeta \rightarrow 0.
\end{equation}

\subsection{Zero surface tension solutions}
\noindent
Let $W_0(\zeta)$ and $z_0(\zeta)$ be the complex potential and conformal map we seek for a bubble without surface tension. We claim that the complex potential for a zero surface tension bubble in an unbounded Hele-Shaw cell is
\begin{equation}
W_0(\zeta)=(1-U)a \left(\zeta + \frac{1}{\zeta} \right).
\label{w0map}
\end{equation}
Taylor \& Saffman \cite{TS} showed that a single steadily translating bubble without surface tension, away from any channel walls, is elliptical in shape. As is shown by Crowdy \cite{Crowdy2009}, the conformal map $z_0(\zeta)$ from $D_\zeta$ to the exterior $D$ of an elliptical bubble centred at $z=0$ is
\begin{equation}
z_0(\zeta)=\frac{a}{\zeta}+a\left(1-\frac{2}{U}\right)\zeta.
\label{z0map}
\end{equation}
Note that this conformal map $z_0(\zeta)$ takes the upper unit $\zeta$-circle to the lower bubble boundary. We note that both $W_0(\zeta)$ and $z_0(\zeta)$ are given in terms of a single real parameter $a$ which is related to the bubble area $A_0$ by
\begin{equation}
A_0=-\frac{1}{2\mathrm{i}}\oint_{|\zeta|=1}\overline{z(\zeta)}z'(\zeta)d\zeta = \frac{4\pi a^2 (U-1)}{U^2}
\end{equation}
where we have used Green's Theorem. The minus sign is necessary to ensure a positive quantity because always $U>1$. Fixing $A_0=\pi$ readily implies that
\begin{equation}
a=\frac{U}{2\sqrt{U-1}}.
\end{equation}
For later reference, we note that for the special case $U=2$, the elliptical bubble reduces to the unit circle via the map $z_0(\zeta)=1/\zeta$.

\subsection{Non-zero surface tension solutions}
\noindent
To incorporate the effect of a non-zero surface tension on the bubble boundary, it therefore seems reasonable to seek a conformal map from $D_\zeta$ to $D$ in the form
\begin{equation}
z(\zeta)=z_0(\zeta)+f(\zeta)
\end{equation}
where $f(\zeta)$ can be viewed as a perturbation function to the conformal mapping $z_0(\zeta)$ in order to suitably compensate for non-zero surface tension effects on the bubble boundary.

\vskip 0.1truein
\noindent
With the inclusion of surface tension on the bubble boundary, we have the following boundary conditions to satisfy:
\begin{align}
\mathrm{Im}[W(\zeta)]&=0, \quad \zeta \in C_0; \label{bc1} \\
\mathrm{Re}[W(\zeta)+Uz(\zeta)]&=B\kappa+\textrm{constant}, \quad \zeta \in C_0; \label{bc2} \\
W(\zeta) &\sim \frac{(1-U)a}{\zeta}, \quad \zeta \rightarrow 0.
\label{bc3}
\end{align}
We claim that the one-parameter family of solutions we need are the following:
\begin{align}
W(\zeta) &\equiv W_0(\zeta), \quad \mathrm{and} \label{ww} \\
z(\zeta) &\equiv z_0(\zeta)+f(\zeta),
\label{zz}
\end{align}
with $W_0(\zeta)$ and $z_0(\zeta)$ given in (\ref{w0map}) and (\ref{z0map}), respectively. Here, $f(\zeta)$ is a perturbation function, analytic everywhere in $D_\zeta$, that incorporates the new physics from the addition of surface tension to the bubble boundary. Note that adding a perturbation function $g(\zeta)$ to the complex potential is not necessary as this will invoke unnecessary calculations (e.g. enforcing $\mathrm{Im}[g(\zeta)]=0$ on $|\zeta|=1$, and would complicate equation (\ref{steqn}) below); all the necessary and sufficient boundary conditions can be enforced using the single perturbation function $f(\zeta)$. This function must behave as follows:
\begin{equation}
f(\zeta) \sim \mathrm{constant}, \quad \zeta \rightarrow 0.
\end{equation}
This constant will set the centroid location of the bubble.

\vskip 0.1truein
\noindent
It is clear that (\ref{bc1}) and (\ref{bc3}) are satisfied by (\ref{ww}) and (\ref{zz}). To ensure that (\ref{bc2}) is satisfied by (\ref{ww}) and (\ref{zz}) also, note that it is a simple exercise to establish
\begin{equation}
\mathrm{Re}[W(\zeta)+U z(\zeta)] \equiv U \mathrm{Re}[f(\zeta)], \quad \zeta \in C_0.
\end{equation}
Thus we are left with the following equation to enforce:
\begin{equation}
U \mathrm{Re}[f(\zeta)]= -B \left(\frac{1+\mathrm{Re}[\zeta \left(z''_0(\zeta)+f''(\zeta)\right)/\left(z'_0(\zeta)+f'(\zeta)\right)]}{|z'_0(\zeta)+f'(\zeta)|} \right), \quad \zeta \in C_0.
\label{steqn}
\end{equation}
That is, to solve the problem, it is left to determine function $f(\zeta)$, along with the two real numbers $a$ and $U$, satisfying equation (\ref{steqn}) for some value of the surface tension parameter $B$ and with the area of the bubble $A = \pi$ fixed:
\begin{equation}
\pi=-\frac{1}{2\mathrm{i}}\oint_{|\zeta|=1}\overline{z(\zeta)}z'(\zeta)d\zeta.
\label{areacond}
\end{equation}
\noindent
Once this has been achieved, we may plot the shape of the bubble free boundary via (\ref{zz}) as this will be the image of $C_0$ under the conformal mapping $z(\zeta)=z_0(\zeta)+f(\zeta)$.

\vskip 0.1truein
\noindent
We note the existence of the following trivial solutions for any non-zero surface tension $B\ne0$ in (\ref{steqn}), satisfying (\ref{areacond}):
\begin{equation}
f(\zeta) = B/2, \quad U=2, \quad a=1.
\label{trivialSol2}
\end{equation}
These solutions, noted by Tanveer~\cite{Tanveer86}, correspond to circular bubbles of unit radius (as can easily be seen from the mapping $z_0(\zeta)$). We refer to this branch of solution as the $m=0$ branch.  Thus we see that the circular solution with $U=2$ is selected as $B\rightarrow 0$ in a trivial way (recall that, of the continuum of elliptic bubbles for $B=0$, the special case $U=2$ is circular).  Motivated by the existence of multiple branches of solution in the related bubble and finger problems in a channel geometry \cite{Combescot,Tanveer86,McLeanSaffman,VDB1983,GardinerEtAlb}, we now look for other solutions, apart from (\ref{trivialSol2}), using numerical techniques.

\subsection{Numerical scheme}
We solve for the perturbation function $f(\zeta)$ numerically by writing it as a truncated Taylor series
\begin{equation}
f(\zeta)=\sum_{j=0}^{N-1} a_j \zeta^{j},
\label{fTaylor}
\end{equation}
and computing the $N$ real coefficients $\{a_j\}_{j=0}^{N-1}$, and the two real parameters $a$ and $U$. We chose $N=200$ for all our calculations. Note that all the coefficients $\{a_j\}_{j=0}^{N-1}$ are indeed necessarily real because the bubble is assumed to be symmetric about the real axis. We have $N+2$ real unknowns and $N+2$ real equations to enforce: (\ref{steqn}) at $N+1$ equi-spaced points on the upper arc of $C_0$, $\{\exp(\pi \mathrm{i} (j-1) / N))\}_{j=1}^{N+1}$, and the bubble area constraint (\ref{areacond}). Thus, the counting is consistent, and we may use the multi-dimensional Newton's method to find solutions. This numerical scheme is well-known to converge quadratically for initial estimates in suitable basins of attraction. It can be shown using Cauchy's residue theorem that, with (\ref{fTaylor}), the area constraint (\ref{areacond}) reduces to the following simple algebraic expression:
\begin{equation}
1=a^2-\left(a_1+a\left(1-\frac{2}{U}\right)\right)^2-\sum_{j=2}^{N-1}j a^{2}_{j}.
\label{algebraiccond}
\end{equation}
Since all lengths in our problem are scaled with respect to a particular bubble area (recall (\ref{scalings})), it is unnecessary to consider different bubble areas. Once a solution is computed for a given value of $B$, a standard continuation procedure in $B$ can be used to trace-out the full branch of solutions.

\begin{figure}[H]
\begin{center}
\includegraphics[scale=0.5,angle=90]{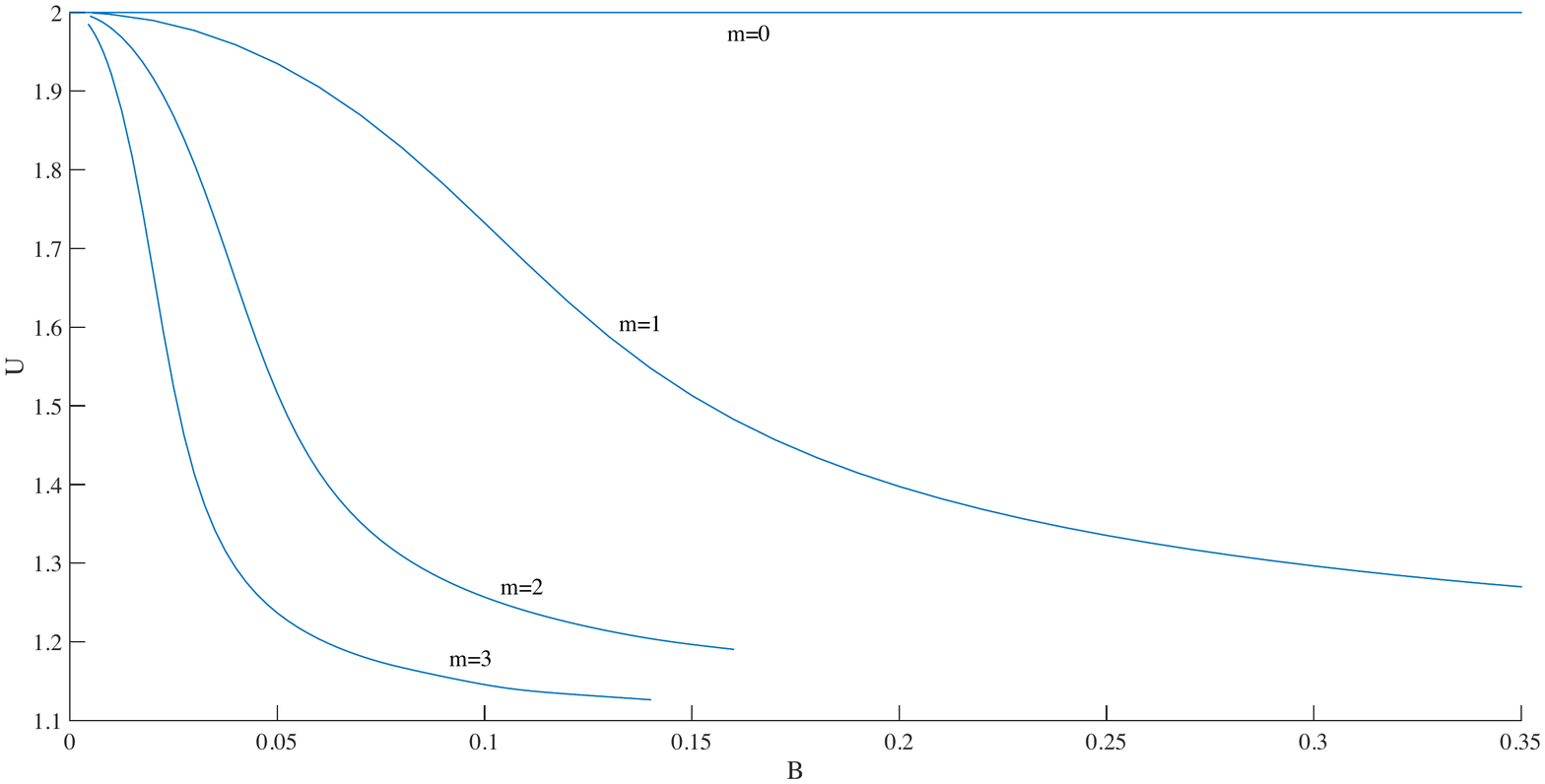}
\vskip -0.8truein
\includegraphics[scale=0.35]{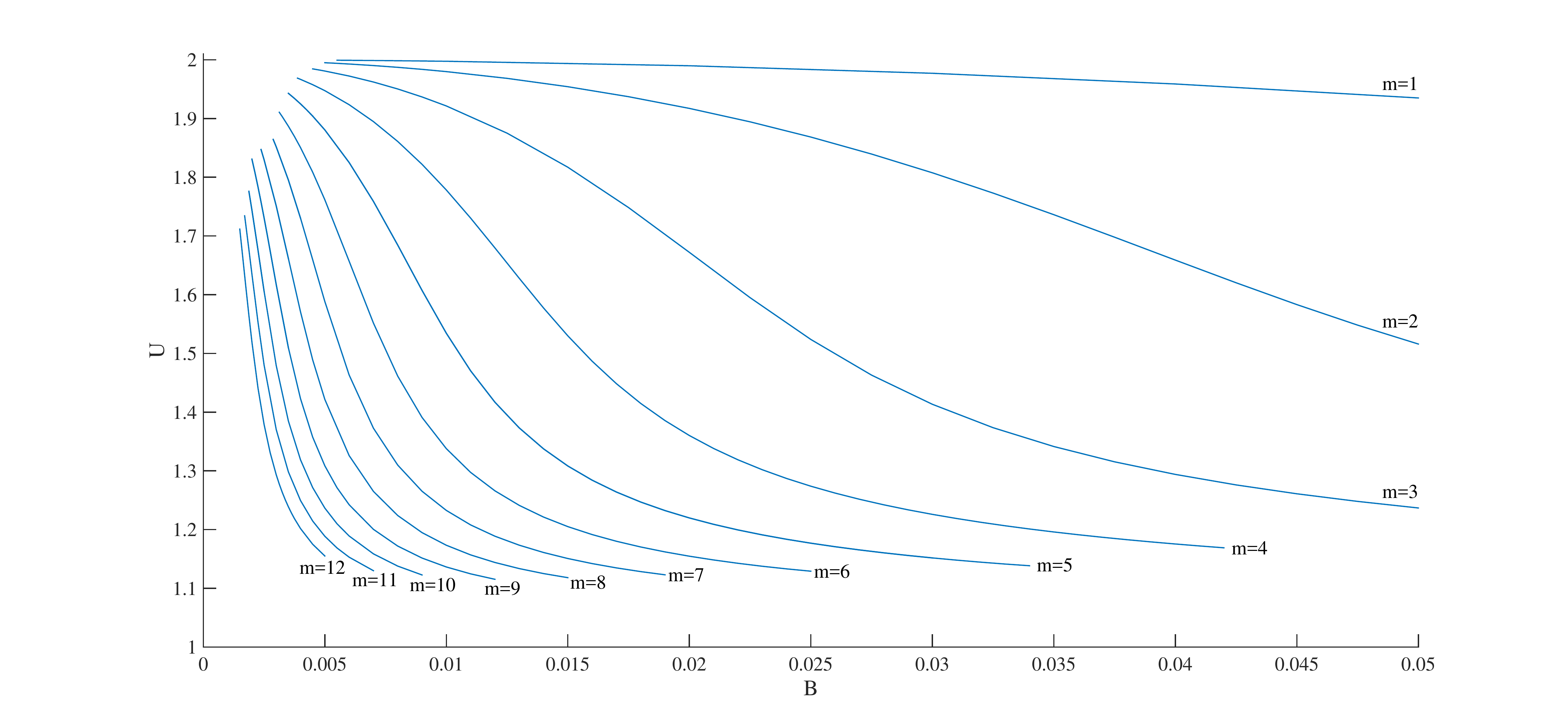}
\caption{Plot of bubble speed $U$ as a function of the surface tension parameter $B$ for (a) solution branches $m=0,1,2,3$ and (b) solution branches $1 \leq m \leq 12$.  Note the trivial solution branch $m=0$ is the horizontal line $U=2$.}
\label{UvB}
\end{center}
\end{figure}

\vskip 0.1truein
\noindent
As mentioned in the previous subsection, the $m=0$ branch of solutions (\ref{trivialSol2}) corresponds to circular bubbles of unit radius.  All of these solutions have a bubble speed $U=2$, and thus we can draw the $m=0$ branch on a $U$ versus $B$ plot as a horizontal line, as in Figure~\ref{UvB}(a).  Interestingly, for any reasonable initial guess (which is not precisely or very close to (\ref{trivialSol2})), our Newton code did not converge to this trivial solution, but instead converged to the next branch of solutions, which we call the $m=1$ branch. By applying a continuation procedure, we computed solutions on this branch, with the dependence of the bubble speed $U$ on the surface tension $B$ shown in Figure~\ref{UvB}(a).  We see that $U$ is monotonically decreasing with $B$, and it appears that $U\rightarrow 2$ as $B\rightarrow 0$.

\begin{figure}[H]
\begin{center}
\includegraphics[scale=0.4]{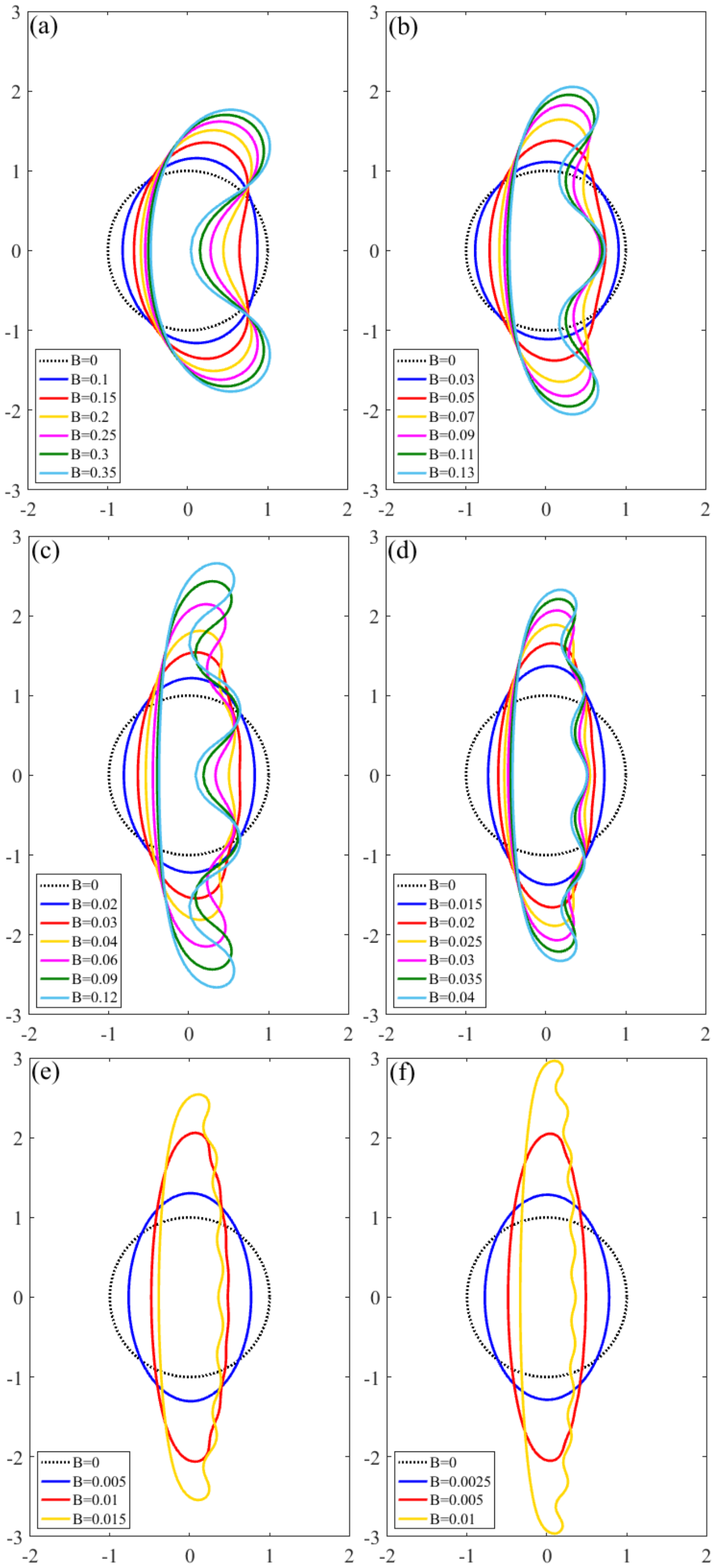}
\caption{Bubble shapes for increasing values of the surface tension parameter $B$: (a) $m=1$ branch; (b) $m=2$ branch; (c) $m=3$ branch; (d) $m=4$ branch; (e) $m=7$ branch; (f) $m=10$ branch.}
\label{plots1}
\end{center}
\end{figure}

\vskip 0.1truein
\noindent
The shape of the bubbles along the $m=1$ branch is interesting, as can be seen in Figure~\ref{plots1}(a).  For small values of surface tension $B$, the bubble is nearly circular.  As $B$ increases, the bubble deforms until it becomes non-convex at a finite value of $B$.  For higher surface tension values, the bubble appears to have two ``tips'' or ``dimples''.  This behaviour is qualitatively similar to that observed by Tanveer~\cite{Tanveer87} for a bubble in a channel geometry; he referred to the analogous branch of solutions as the `extraordinary' branch.  Further, these shapes are similar to the double-tipped finger solutions computed by Franco-G\'omez et al. \cite{FrancoGomez2016}, Gardiner et al.\cite{GardinerEtAlb} and Thompson et al.~\cite{Thompson2014}.

\begin{figure}
\begin{center}
\includegraphics[scale=0.5,angle=90]{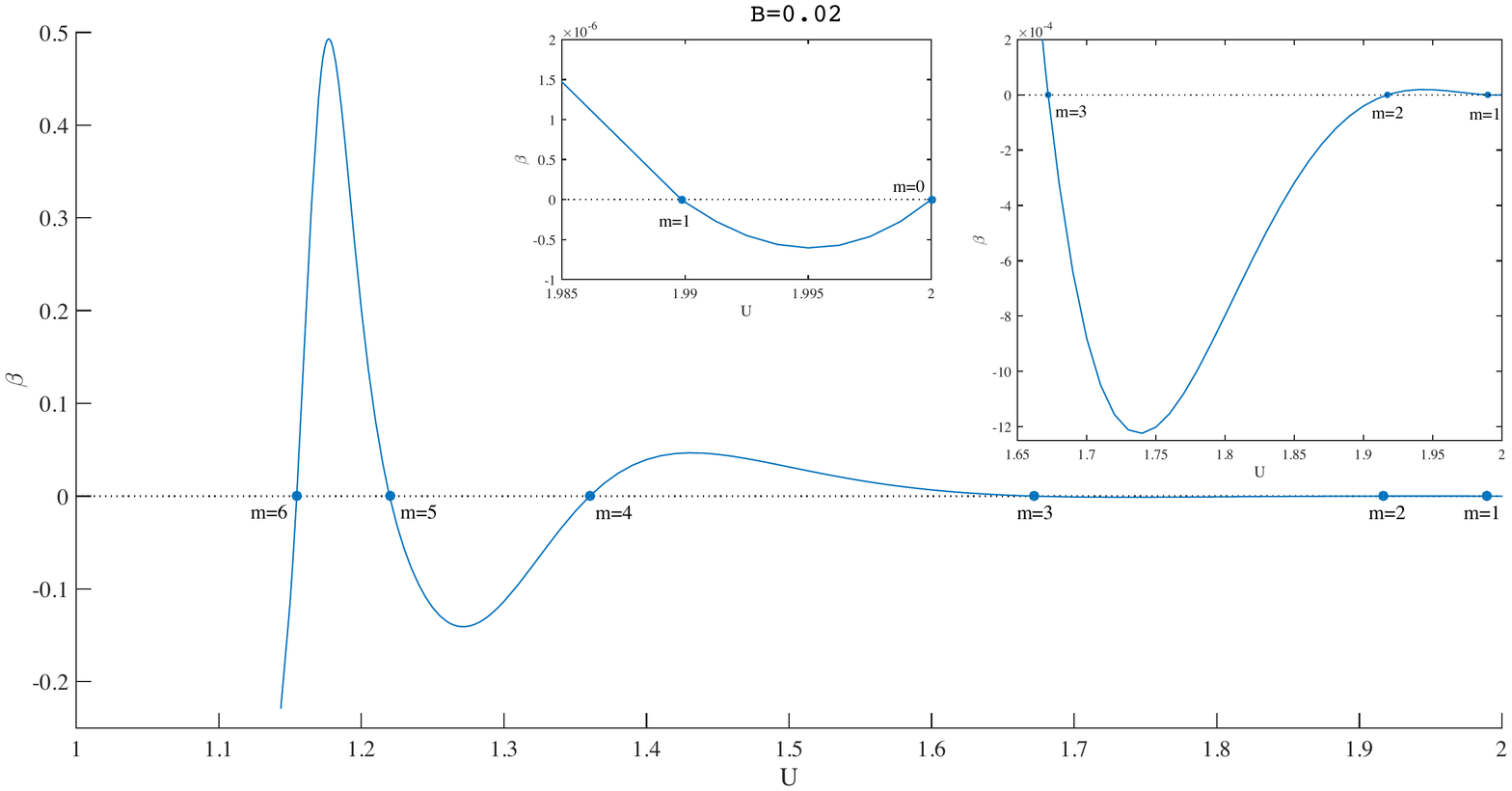}
\vskip -0.75truein
\caption{Plot of $\beta$ as a function of $U$ for $B=0.02$ fixed. Shown by dots are the seven bubble solutions found corresponding to the branches $m=0,1,2,...6$ for this value of the surface tension.}
\label{beta1}
\end{center}
\end{figure}

\vskip 0.1truein
\noindent
Locking on to other solution branches for $m\geq 2$ was done using the following technique. We introduce a new parameter $\beta$ such that
\begin{equation}
U \mathrm{Re}[f(1)]=-B \left(\frac{1+\mathrm{Re}[\left(z''_0(1)+f''(1)\right)/\left(z'_0(1)+f'(1)\right)]}{|z'_0(1)+f'(1)|} \right)+\beta.
\label{betaEqn}
\end{equation}
We then solve (\ref{betaEqn}) together with (\ref{steqn}) at the remaining points along the upper unit $\zeta$-circle for $\beta$, $a$ and $\{a_j\}_{j=0}^{N-1}$ for given fixed values of $U$ and $B$. The physical solution space then corresponds to solutions with $\beta=0$ (for $\beta \ne 0$, the artificial solution space corresponds to bubbles with a corner/sharp tip at the leading edge). This approach was adopted by \cite{GardinerEtAlb}, for instance, for the finger problem. Figure~\ref{beta1} shows a plot of $\beta$ as a function of $U$ for $B=0.02$ fixed. The points of intersection with the line $\beta=0$ indicate the physical bubble solutions. All solutions for $\beta$ as a function of $U$ are shown. This is corroborated by the data in Figure \ref{UvB}(b) which reveals the existence of six solution branches having $B=0.02$ (in addition to the trivial solution for $m=0$).

\begin{figure}
\begin{center}
\includegraphics[scale=0.5,angle=90]{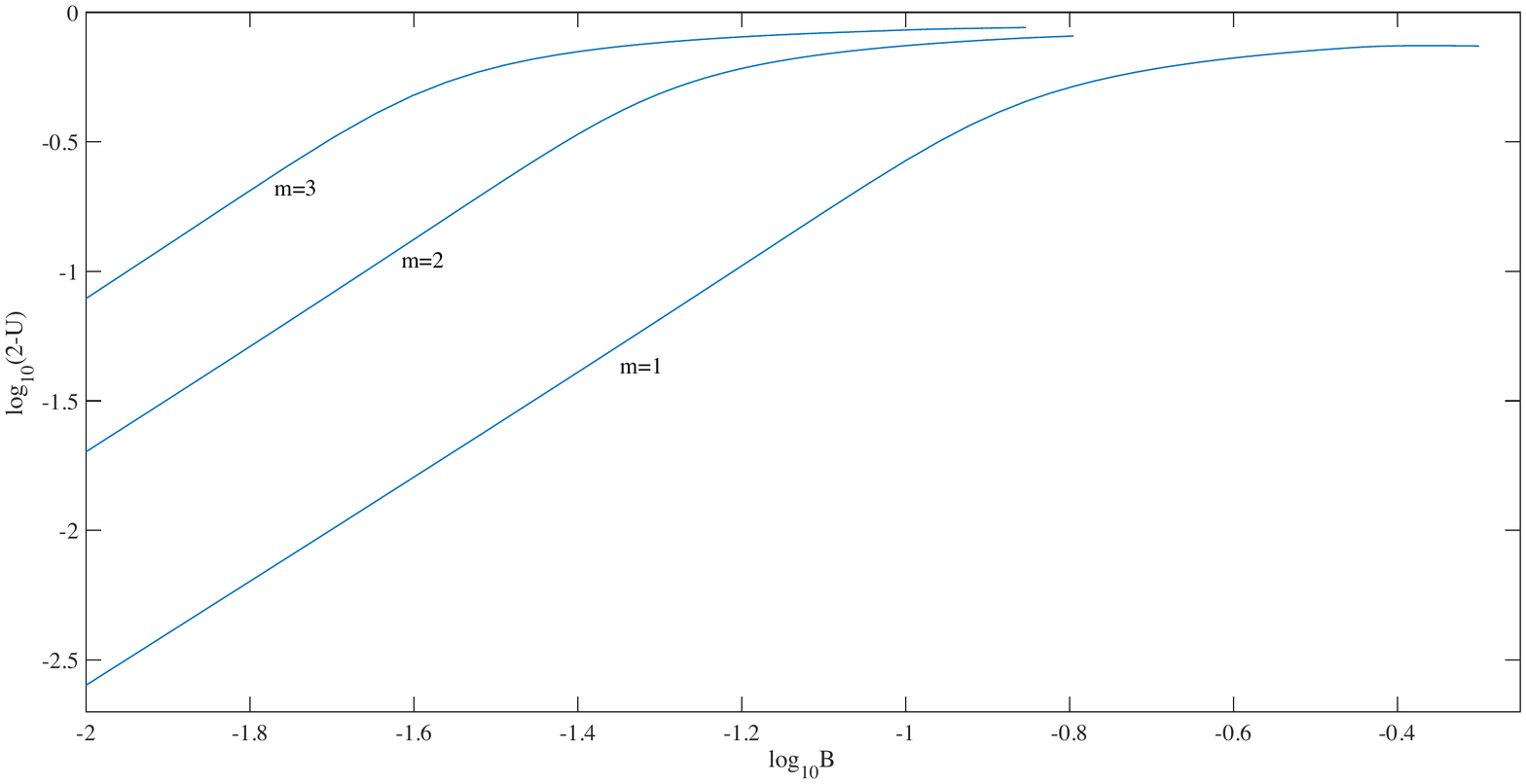}
\vskip -0.75truein
\caption{Plot of $\log_{10}(2-U)$ versus $\log_{10}B$ for solution branches $m=1,2,3$.}
\label{loglog}
\end{center}
\end{figure}

\vskip 0.1truein
\noindent
Our results for $U$ versus $B$ for solution branches up to $m=12$ are shown in Figure~\ref{UvB}.  It appears from this data that $U\rightarrow 2$ as $B\rightarrow 0$ on all of the branches, which is consistent with the previously mentioned studies for a steadily propagating bubble in a channel geometry~\cite{Combescot,Tanveer86,Tanveer87,Tanveer89} (as well as the analogous problem for a Saffman-Taylor finger propagating in a channel \cite{McLeanSaffman,VDB1983,GardinerEtAlb,CombescotEtAl,Chapman99,Tanveer872,Hong,Shraiman}).  In Figure~\ref{loglog} we present the same data for the first few non-trivial branches, but this time on a log-log plot.  We see that as $\log B \rightarrow -\infty$, the data appears to follow a straight line with slope 2, which suggests that
\begin{equation}
U \sim 2-kB^{2}, \quad B \rightarrow 0,
\end{equation}
where $k$ is a constant that depends on $m$.  This quadratic scaling is interesting because it is different to the analogous bubble and finger problems in a channel, for which the relevant scaling is $U\sim 2-kB^{2/3}$.  Note that this finding is not inconsistent with the results for a bubble in a channel geometry, because the limit of the channel walls moving to infinity does \textit{not} commute with the limit of the vanishing surface tension for a fixed channel. (It is also worth noting that the limit of the channel walls tending to infinity for given surface tension is not equivalent to the limit of bubble area tending to zero in a channel with fixed wall separation).

\vskip 0.1truein
\noindent Returning to the shape of the bubbles in Figure~\ref{plots1}, each branch of solutions follows a certain pattern.  We have already noted that the  $m=0$ solutions are all circles, which can be thought of as having a single tip, and the $m=1$ solutions have a double tip (for sufficiently large $B$).  In Figure~\ref{plots1}(b), we see for the $m=2$ branch that for small surface tension the bubbles are again convex; however, for sufficiently large surface tension, the bubbles become non-convex with three tips.  The pattern continues in Figures~\ref{plots1}(c), (d), etc., so that we observe bubble shapes with $m+1$ tips on the $m$-th branch of solutions.  Further, the right-most point at which the bubble intersects the $x$-axis alternates from being locally convex ($m$ even) to locally concave ($m$ odd) as $m$ increases.  This multiple tip and alternating convex/concave behaviour is analogous to the behaviour of Saffman-Taylor fingers \cite{GardinerEtAlb}.

\section{Two bubbles}
\noindent
The selection mechanism for more than one bubble does not appear to have received any prior attention. To make progress, we will consider the special case of two up-down symmetric bubbles in steady motion in the unbounded Hele-Shaw cell. This problem presents several interesting key features, not least because the separation between the two bubbles is now another key factor in the selection mechanism.   Crucially, this symmetric pair of bubbles also has a zero surface tension solution governing their shapes (Crowdy \cite{Crowdy2009}), as we discuss later in this section (see (\ref{W0mapp})-(\ref{z0mapp})).

\subsection{Problem formulation}
\noindent
Consider two bubbles which are up-down symmetric in shape (by which we mean symmetric through the real axis, say) in steady motion in an unbounded Hele-Shaw cell. The up-down symmetry of this bubble configuration affords us several pleasant analytical simplifications. Let $D$ be the unbounded planar region, in the $z=x+\mathrm{i}y$-plane, containing incompressible fluid exterior to the bubble pair, with each bubble having the same finite area. Denote the upper bubble by $D_0$ and its boundary by $\partial D_0$; denote the lower bubble by $D_1$ and its boundary by $\partial D_1$. We will assume that the bubble pair is moving with constant speed $U$ in the $x$-direction, and that the fluid velocity far away from the bubbles is in the $x$-direction with constant speed $V$. The velocity field $\mathbf{u}$ is derived from a velocity potential $\phi$ so that $\mathbf{u}=\nabla \phi$.

\begin{figure}
\begin{center}
\includegraphics[scale=0.3]{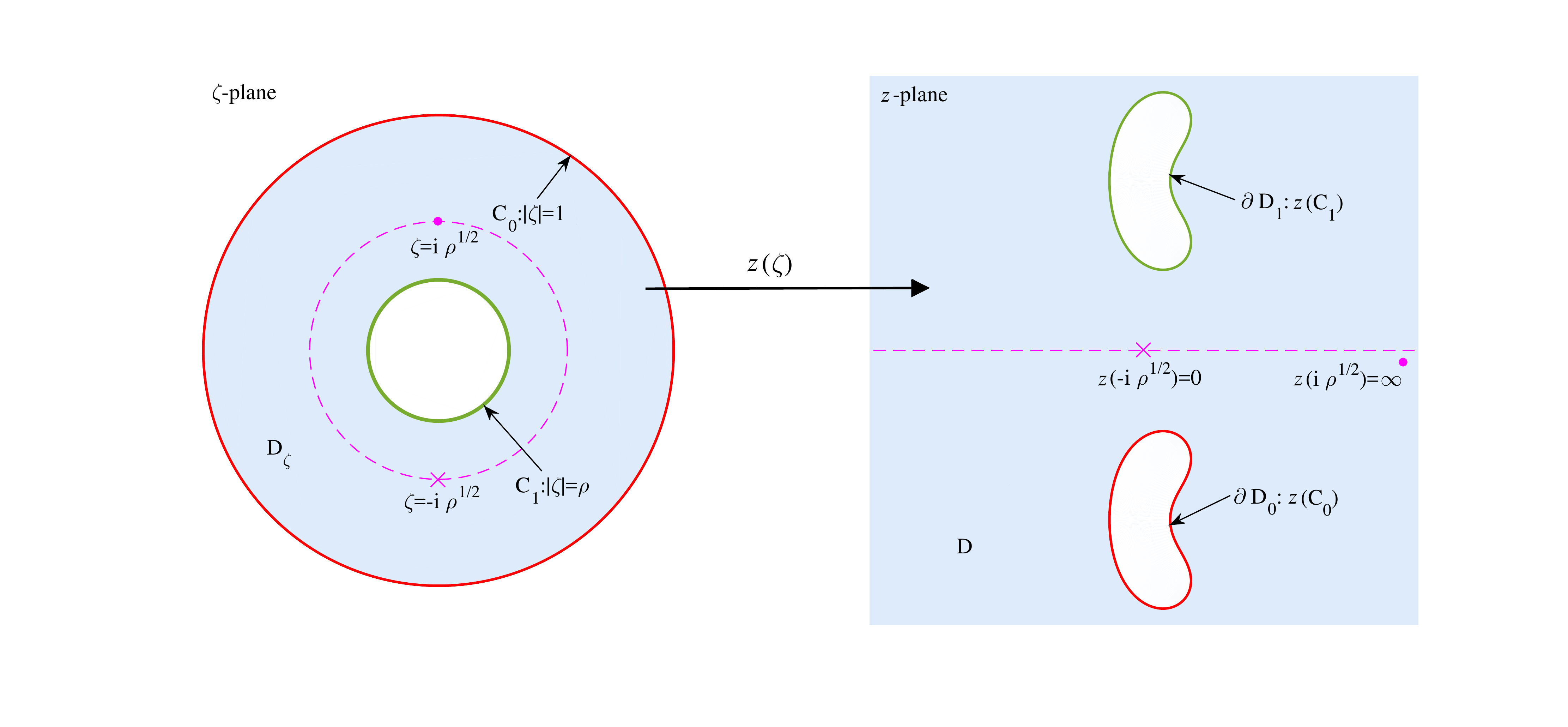}
\vskip -0.3truein
\caption{Schematic of the pre-image parametric $\zeta$-plane being the interior of the annulus $\rho<|\zeta|<1$ with boundaries consisting of the two circles $C_0$ and $C_1$ (left), and the target image domain in the $z$-plane exterior to the two bubbles (right).}
\label{schematic2}
\end{center}
\end{figure}

\vskip 0.1truein
\noindent
We thus have the following problem to solve, for the velocity potential $\phi$ and the streamfunction $\psi$, pertaining to a two up-down symmetric bubbles in an unbounded Hele-Shaw cell in a frame of reference co-travelling with the bubble pair at speed $U$:
\begin{align}
\nabla^2 \phi &=0, \quad z \in D; \label{goveq11} \\
\phi + U x &= B \kappa + \phi_0, \quad z \in \partial D_0; \label{goveq22} \\
\phi + U x &= B \kappa + \phi_1, \quad z \in \partial D_1; \label{goveq33} \\
\psi &= 0, \quad z \in \partial D_0; \label{goveq44} \\
\psi &= \psi_1, \quad z \in \partial D_1; \label{goveq55} \\
\phi &\sim (1-U) x, \quad |z|\rightarrow \infty. \label{goveq66}
\end{align}
In the above, $\phi_0$, $\phi_1$ and $\psi_1$ are real constants. It makes sense, thus, to solve for a complex potential function $w(z)=\phi+\mathrm{i}\psi$ for this flow whose real and imaginary parts are respectively the velocity potential $\phi$ and the streamfunction $\psi$. Function $w$ must be a single-valued analytic function everywhere in $D$. A natural way of doing this is to pull-back to a simpler parametric $\zeta$-plane and proceed to find the composition $W(\zeta)=w(z(\zeta))$, where $z(\zeta)$ is a conformal map to the fluid region $D$ exterior to the bubble, and must also be determined. Label the pre-image region of $D$ in the $\zeta$-plane by $D_\zeta$ which may be taken to be the concentric annular region $\rho<|\zeta|<1$, without loss of generality. Let $C_0$ label the circle $|\zeta|=1$ and $C_1$ label the circle $|\zeta|=\rho$. The Riemann-Koebe mapping theorem \cite{Goluzin} guarantees the existence of a conformal mapping $z(\zeta)$ between the two doubly connected regions $D_\zeta$ and $D$. We will now re-write the problem (\ref{goveq11})-(\ref{goveq66}) in terms of the complex functions $z(\zeta)$ and $W(\zeta)$, or equivalently, in terms of the complex variable $\zeta$. Figure \ref{schematic2} shows a schematic.

\vskip 0.1truein
\noindent
In the frame of reference co-travelling with the two bubbles, the complex potential $w(z)$ is related to the complex potential $\tilde w(z)$ in the laboratory frame via $w(z)=\tilde w(z)-Uz$. It follows that $w$ is a single-valued analytic function everywhere in $D$ except for a simple pole singularity at infinity. Equivalently, $W(\zeta)$ is a single-valued analytic function everywhere in $D_\zeta$ except for a simple pole singularity at the point $\zeta=\beta \in D_\zeta$ (say) mapping to infinity. Correspondingly, the conformal map $z(\zeta)$ must have a simple pole also at $\zeta=\beta$ but is otherwise analytic and single-valued. This means that locally, we must have
\begin{equation}
z(\zeta) \sim \frac{a}{\zeta-\beta}, \quad \zeta \rightarrow \beta,
\label{zbehavv}
\end{equation}
for some constant $a$, which can be taken to be real; this leaves two remaining real degrees of freedom associated with the Riemann mapping theorem which allows us to choose $\beta=\mathrm{i}\sqrt{\rho}$. We will construct $z(\zeta)$ such that $C_1$ maps to the upper bubble boundary $\partial D_0$ and $C_0$ maps to the lower bubble boundary $\partial D_1$.

\vskip 0.1truein
\noindent
It is a trivial observation that if $W(\zeta)=\Phi + \mathrm{i} \Psi$ is a single-valued analytic function of $\zeta$, then both its real and imaginary parts $\Phi$ and $\Psi$ are necessarily harmonic, as required.
(\ref{goveq22}) and (\ref{goveq33}) become
\begin{align}
\mathrm{Re}[W(\zeta)+Uz(\zeta)]&=B\kappa+c_0, \quad \zeta \in C_0, \quad \mathrm{and} \\
\mathrm{Re}[W(\zeta)+Uz(\zeta)]&=B\kappa+c_1, \quad \zeta \in C_1,
\end{align}
where an expression for the curvature $\kappa$ is as in (\ref{curvature}), and $c_0$ and $c_1$ are real constants. (\ref{goveq44}) and (\ref{goveq55}) are simply
\begin{align}
\mathrm{Im}[W(\zeta)]&=0, \quad \zeta \in C_0, \quad \mathrm{and} \\
\mathrm{Im}[W(\zeta)]&=d_1, \quad \zeta \in C_1,
\end{align}
where $d_1$ is a real constant. (\ref{goveq66}) transforms to
\begin{equation}
W(\zeta) \sim (1-U)z(\zeta), \quad \zeta \rightarrow \beta,
\end{equation}
which, in light of (\ref{zbehavv}) and the choice $\beta=\mathrm{i}\sqrt{\rho}$, is
\begin{equation}
W(\zeta) \sim \frac{(1-U)a}{\zeta-\mathrm{i}\sqrt{\rho}}, \quad \zeta \rightarrow \mathrm{i}\sqrt{\rho}.
\end{equation}

\subsection{Zero surface tension solutions}
\noindent
Let $W_0(\zeta)$ and $z_0(\zeta)$ be the complex potential and conformal map we seek for the bubble pair without surface tension. The complex potential for a pair of up-down symmetric bubbles with zero surface tension in an unbounded Hele-Shaw cell can be derived using the most general result (for any finite number of bubbles) in Crowdy \cite{Crowdy2009}:
\begin{equation}
W_0(\zeta)=\frac{\mathrm{i}a(1-U)}{\sqrt{\rho}} \left[1-K(-\mathrm{i}\zeta/\sqrt{\rho};\rho)-K(-\mathrm{i}\zeta \sqrt{\rho};\rho)\right].
\label{W0mapp}
\end{equation}
Using the other general result of \cite{Crowdy2009}, the conformal map $z_0(\zeta)$ from $D_\zeta$ to the exterior $D$ of an up-down symmetric pair of bubbles (about the real axis) is
\begin{equation}
z_0(\zeta)=\frac{\mathrm{i}a}{\sqrt{\rho}}\left[\frac{1}{2}-K(-\mathrm{i}\zeta /\sqrt{\rho};\rho)-\left(1-\frac{2}{U}\right)K(-\mathrm{i}\zeta \sqrt{\rho};\rho)\right].
\label{z0mapp}
\end{equation}
Here, the function $K(\zeta;\rho)$ is defined through the following logarithmic derivative relation
\begin{equation}
K(\zeta;\rho)=\zeta \frac{d}{d\zeta} \log P(\zeta;\rho)=\frac{\zeta}{\zeta-1}-\sum_{j=1}^{\infty}\frac{\rho^{2j}\zeta}{1-\rho^{2j}\zeta}+\sum_{j=1}^{\infty}\frac{\rho^{2j}\zeta^{-1}}{1-\rho^{2j}\zeta^{-1}},
\label{K}
\end{equation}
where $P(\zeta;\rho)$ is the Schottky-Klein prime function associated with $D_\zeta$ and is given by the rapidly convergent infinite product
\begin{equation}
P(\zeta;\rho)=(1-\zeta) \prod_{j=1}^{\infty}(1-\rho^{2j}\zeta)(1-\rho^{2j}\zeta^{-1}).
\label{P}
\end{equation}
For an overview of the Schottky-Klein prime function and a novel way to compute it, see \cite{Crowdy2008,computSK}.  The $P$-function (\ref{P}) arises naturally when mapping from an annulus and has been used in the analysis of other Hele-Shaw flow problems with doubly connected domains~\cite{Crowdy2004,Dallaston2012,Marshall2015,Silva2013}.  Note that the conformal map $z_0(\zeta)$ in (\ref{z0mapp}) maps $C_0$ to the lower bubble and $C_1$ to the upper bubble, the circle $|\zeta|=\sqrt{\rho}$ maps to the real $z$-axis, and $z_0(-\mathrm{i}\sqrt{\rho})=0$.

\vskip 0.1truein
\noindent
We note that both $W_0(\zeta)$ and $z_0(\zeta)$ are given in terms of two real parameters $a$ and $\rho$ which are respectively related to the area of the bubbles $A_0$ by
\begin{equation}
A_0=-\frac{1}{2\mathrm{i}}\oint_{|\zeta|=1}\overline{z(\zeta)}z'(\zeta)d\zeta =-\frac{1}{2\mathrm{i}}\oint_{|\zeta|=\rho}\overline{z(\zeta)}z'(\zeta)d\zeta
\end{equation}
and the separation of the bubbles (the bubbles are closest as $\rho \rightarrow 1$). We may set $A_0=\pi$ as before.

\subsection{Non-zero surface tension solutions}
\noindent
To incorporate the effect of a non-zero surface tension on the bubble boundaries, it therefore seems reasonable to seek a conformal map from $D_\zeta$ to $D$ in the form
\begin{equation}
z(\zeta)=z_0(\zeta)+f(\zeta)
\end{equation}
where $f(\zeta)$ can be viewed as a perturbation function to the conformal mapping $z_0(\zeta)$ in order to suitably compensate for non-zero surface tension effects on the two bubble boundaries.

\vskip 0.1truein
\noindent
With the inclusion of surface tension on the bubble boundary, we have the following boundary conditions to satisfy:
\begin{align}
\mathrm{Im}[W(\zeta)]&=0, \quad \zeta \in C_0; \label{bc111} \\
\mathrm{Im}[W(\zeta)]&=d_1, \quad \zeta \in C_1; \label{bc222} \\
\mathrm{Re}[W(\zeta)+Uz(\zeta)]&=B\kappa+c_0, \quad \zeta \in C_0; \label{bc333} \\
\mathrm{Re}[W(\zeta)+Uz(\zeta)]&=B\kappa+c_1, \quad \zeta \in C_1; \label{bc444} \\
W(\zeta) &\sim \frac{(1-U)a}{\zeta-\mathrm{i}\sqrt{\rho}}, \quad \zeta \rightarrow \mathrm{i}\sqrt{\rho}.\label{bc555}
\end{align}
We claim that the two-parameter family of solutions we need are the following:
\begin{align}
W(\zeta) &\equiv W_0(\zeta), \quad \mathrm{and} \label{www} \\
z(\zeta) &\equiv z_0(\zeta)+f(\zeta),
\label{zzz}
\end{align}
with $W_0(\zeta)$ and $z_0(\zeta)$ given in (\ref{W0mapp}) and (\ref{z0mapp}), respectively. Here, $f(\zeta)$ is a perturbation function, analytic everywhere in the annulus $D_\zeta$. As before, note that adding a perturbation function to the complex potential is not necessary as this will invoke unnecessary calculations. This function must behave as follows:
\begin{equation}
f(\zeta) \sim \mathrm{constant}, \quad \zeta \rightarrow \mathrm{i}\sqrt{\rho}.
\end{equation}
This constant sets the centroid locations of the bubbles.

\vskip 0.1truein
\noindent
It is clear that (\ref{bc111}), (\ref{bc222}) and (\ref{bc555}) are satisfied by (\ref{www}) and (\ref{zzz}). To ensure that (\ref{bc333}) and (\ref{bc444}) are satisfied by (\ref{www}) and (\ref{zzz}) also, note that
\begin{equation}
\mathrm{Re}[W(\zeta)+U z(\zeta)] \equiv U \mathrm{Re}[f(\zeta)], \quad \zeta \in C_0, C_1.
\label{URefc}
\end{equation}
This result is established in the appendix. Thus we are left with the following two equation to enforce:
\begin{equation}
U\mathrm{Re}[f(\zeta)]=-B \left(\frac{1+\mathrm{Re}[\zeta \left(z''_0(\zeta)+f''(\zeta)\right)/\left(z'_0(\zeta)+f'(\zeta)\right)]}{|z'_0(\zeta)+f'(\zeta)|} \right), \quad \zeta \in C_0.
\label{steqn11}
\end{equation}
It is automatic from the up-down symmetry of the bubble configuration that if (\ref{steqn11}) is satisfied for $\zeta \in C_0$, then
\begin{equation}
U\mathrm{Re}[f(\zeta)]=-B \left(\frac{1+\mathrm{Re}[\zeta \left(z''_0(\zeta)+f''(\zeta)\right)/\left(z'_0(\zeta)+f'(\zeta)\right)]}{\rho|z'_0(\zeta)+f'(\zeta)|} \right), \quad \zeta \in C_1.
\end{equation}
That is, to solve the problem, it is left to determine function $f(\zeta)$ satisfying equation (\ref{steqn11}), for a given separation ($\rho$ fixed), as the surface tension parameter $B$ varies and the area of the bubbles $A = \pi$ is fixed:
\begin{equation}
\pi=-\frac{1}{2\mathrm{i}}\oint_{|\zeta|=1}\overline{z(\zeta)}z'(\zeta)d\zeta.
\label{areacond2}
\end{equation}
There is no pleasant way to simply the evaluation of this area integral and produce an algebraic expression for the area (analogous to (\ref{algebraiccond})), so we resort to using the trapezium rule to enforce (\ref{areacond2}) which is well-known to be exponentially accurate for periodic functions.

\vskip 0.1truein
\noindent
To completely solve the problem, we must solve (\ref{steqn11}) with the conformal mapping function $z_0(\zeta)$ given by (\ref{z0mapp}) with the parameters $a$ and $\rho$ fixed by enforcing condition (\ref{areacond2}). Our task is to determine the function $f(\zeta)$ and the number $U$ for some value of $B$. Once this is achieved, we may plot the shape of the free boundaries of the bubbles via (\ref{zzz}) as they will be the images of $C_0$ and $C_1$ under the conformal mapping $z(\zeta)=z_0(\zeta)+f(\zeta)$.

\subsection{Numerical scheme}
\noindent
One may solve for the perturbation function $f(\zeta)$ numerically by writing it as a truncated Laurent series
\begin{equation}
f(\zeta)=\sum_{j=-N}^{N} a_j \zeta^{j},
\label{fLaurent}
\end{equation}
where the $2N+1$ coefficients $\{a_j\}_{j=-N}^{N}$ are in general complex numbers. However, the up-down symmetry of the bubble configuration affords us a simplification through the following analytic relation which must hold for all $\zeta$:
\begin{equation}
z(\rho \zeta)=\overline{z}(1/\zeta).
\end{equation}
It can be verified that map $z_0(\zeta)$ automatically satisfies this relation. We thus require $f(\zeta)$ to do so too; it may be deduced that
\begin{equation}
a_{-j}=\rho^{j}\overline{a_j}, \quad j=1,...,N,
\end{equation}
and $a_0$ is purely real. Hence
\begin{equation}
f(\zeta)=a_0 + \sum_{j=1}^{N} \left(a_j \zeta^{j} + \rho^{j}\overline{a_j} \zeta^{-j} \right).
\label{fLaurent2}
\end{equation}
As before, we took $N=200$. Writing $a_j=a_{jr}+\mathrm{i}a_{ji}$, $j=1,...,N$, we see that we are left to solve for the $2N+1$ real numbers $a_0$, $\{a_{jr}, a_{ji}\}_{j=1}^{N}$, in addition to the two real parameters $a$ and $U$ appearing in $W_0(\zeta)$ and $z_0(\zeta)$, for some fixed value of $\rho$. We have $2N+3$ real unknowns and $2N+3$ real equations to enforce: (\ref{steqn11}) at $2N+2$ equi-spaced points around $C_0$, $\{\exp(2 \pi \mathrm{i} (j-1) / (2N+2)))\}_{j=1}^{2N+2}$, and the bubble area constraint (\ref{areacond2}). Thus, the counting is consistent, and we may use a multi-dimensional Newton's method to find solutions. For a fixed separation $\rho$, once a solution was found for a given value of $B$, a standard continuation procedure in $B$ is used to trace-out the full branch of solutions.

\vskip 0.1truein
\noindent
To solve our system, a Jacobian-free Newton-Krylov method \cite{Knoll} implemented by the SUNDIALS software package KINSOL \cite{Hindmarsh} was employed to reduce computational times (i.e. eliminate the need to re-calculate the Jacobian matrix in each iteration) and allow for the solution of a large number of modes $N$ in (\ref{fLaurent2}); indeed, the retention of late order modes becomes important as $B$ and $\rho$ increase. Using a Jacobian-free Newton-Krylov method was unnecessary in the case of one bubble because the functions involved were far less computationally intensive and the parameter space was smaller.

\vskip 0.1truein
\noindent
An initial observation is that our Newton code did not converge to a solution on a branch that could be considered an analogue of the $m=0$ branch for the single bubble problem, (\ref{trivialSol2}). Thus we do not have evidence that there is a non-trivial version of (\ref{trivialSol2}) for two bubbles.

\begin{figure}
\begin{center}
\includegraphics[scale=0.375]{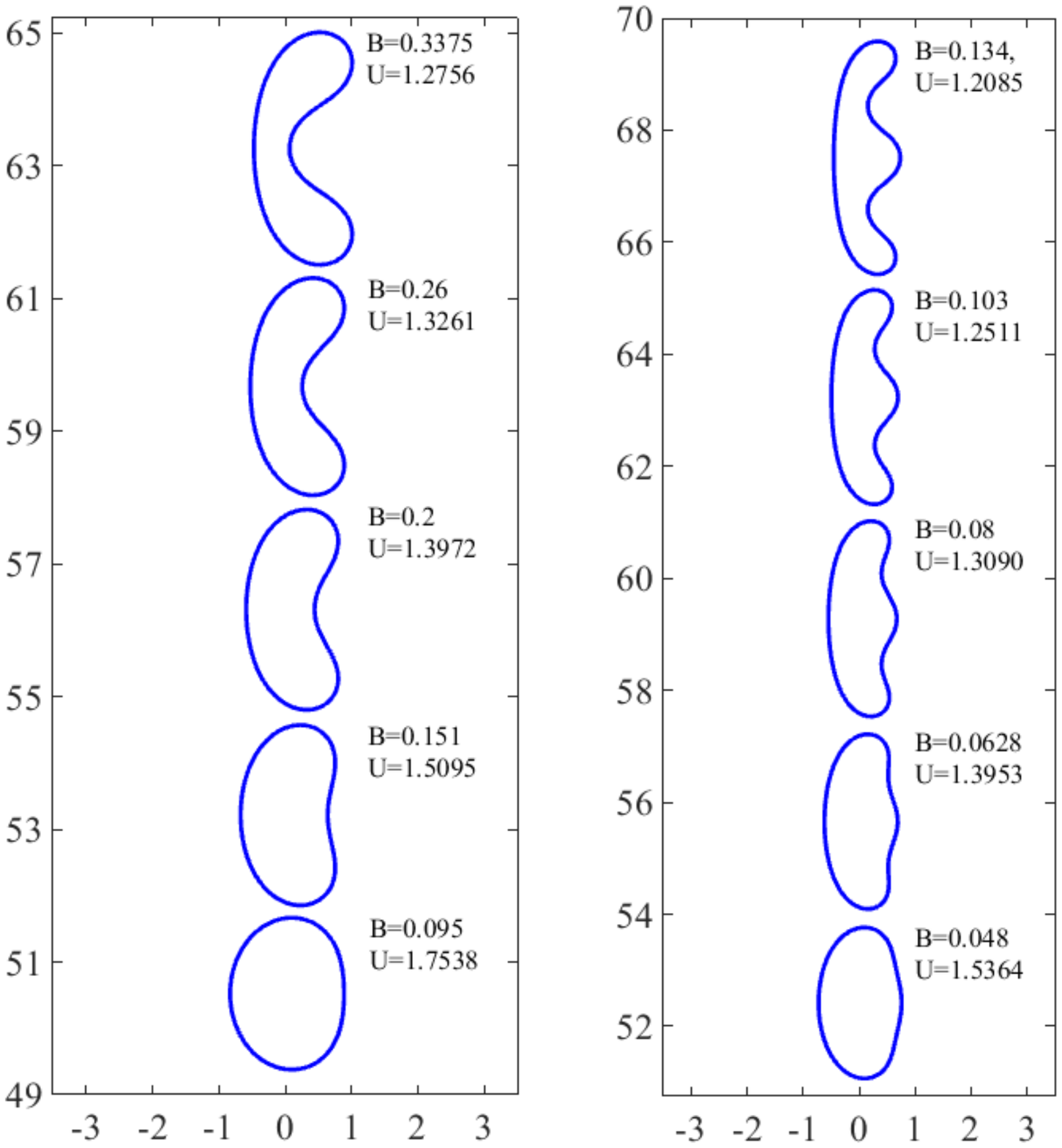}
\vskip -0.8truein
\caption{Upper bubble shapes (lower bubble shapes are identical) for specified values of the surface tension parameter $B$ with the separation parameter $\rho=0.0001$ fixed: $m=1$ branch (left); $m=2$ branch (right).}
\label{2bubfigs}
\end{center}
\end{figure}

\vskip 0.1truein
\noindent
We could, however, compute solutions on the other branches $m\geq 1$. For example, figure \ref{2bubfigs} shows bubbles, each with a different surface tension parameter $B$, on the $m=1$ and $m=2$ solution branches in the upper-half $z$-plane for a fixed separation parameter value $\rho=0.0001$ (the corresponding bubble in the lower-half $z$-plane will assume the same shape, owing to the enforced up-down symmetry). This value of $\rho$ is quite small, so the two bubbles will be far apart. We see that as $B$ increases, the bubble shapes appear to be qualitatively very similar to those that we computed using the simply connected mapping in the previous section; this is reassuring given that the functions involved with the doubly connected mapping in this section are indeed non-trivial. For these large separations, the speed $U$ selected and the Taylor series part of the perturbation function are very close to the analogous solutions for one bubble. As can also be seen in Figure \ref{2bubfigs}, with the surface tension parameter reduced to zero, the shape of the bubbles on both solution branches returns to the shape determined by the zero surface tension conformal map which, given that these bubbles are well-separated, are shapes which near circular.

\vskip 0.1truein
\noindent
We were not able to generate solutions of sufficient accuracy as $\rho \sim \mathcal{O}(0.1)$ which we suspect might be due to the late order modes in (\ref{fLaurent2}) becoming increasingly challenging to compute sufficiently accurately as both $B$ and $\rho$ increase. It is expected that the exotic shapes we observed for the single bubbles will be replicated very closely on each of the solution branches $m\geq 1$ when the two bubbles are well-separated (small $\rho$). As the bubbles become closer together (larger $\rho$), for a fixed value of $B$, further interesting shapes are expected as the bubbles begin to interact.

\section{Discussion}

In this paper, we consider steadily propagating bubbles in an unbounded Hele-Shaw cell. We apply conformal mapping and numerical techniques to compute their shapes and their speed as a function of the dimensionless surface tension parameter $B$. The important idea is to write the solution for the conformal map as the sum of the zero surface tension solution $z_0(\zeta)$ and an unknown analytic function $f(\zeta)$ whose real part satisfies a highly non-linear condition on the unit circle in the conformally mapped $\zeta$-plane. We consider two geometries. For a single propagating bubble, the zero surface tension solution $z_0(\zeta)$ is associated with a simply connected map from the unit disc in the $\zeta$-plane to the outside of the bubble in the physical plane; the straightforward solution represents a one-parameter family of elliptic bubbles where $U$ is the parameter. For two bubbles that are up-down symmetric, the doubly connected solution involves mapping from an annulus in the $\zeta$-plane. Here, the zero surface tension solution $z_0(\zeta)$ is due to Crowdy \cite{Crowdy2009} and is related to the Schottky-Klein prime function \cite{Crowdy2008,computSK}.  This two-bubble solution depends continuously on two parameters, the bubble speed $U$ and a mapping parameter $\rho$ that has the property that $\rho\rightarrow 0$ as the separation distance between the two bubbles goes to infinity.

For both the single bubble and two-bubble problems, our numerical results demonstrate a countably infinite number of solution branches for all $B>0$, which we label $m=0,1,2,\ldots$.
For each solution branch, the bubble speed is $U<2$, with the selection of the $U=2$ solution in the limit $B\rightarrow 0$. The existence of a discrete set of solutions and the selection of $U=2$ in the limit $B\rightarrow 0$ is perfectly analogous to the related bubble and finger problems in a channel geometry \cite{Combescot,Tanveer86,McLeanSaffman,VDB1983,GardinerEtAlb}. However, it is interesting to note that our numerical results suggest that the scaling for our unbounded Hele-Shaw cell is $U\sim 2-kB^2$ as $B\rightarrow 0$, which is different to the scaling for the channel problems, namely $U\sim 2-kB^{2/3}$ as $B\rightarrow 0$.  A worthwhile exercise would be to confirm this new scaling by analysing the selection problem analytically using exponential asymptotics.

Another interesting feature of our numerical results is that for sufficiently large surface tension, the bubbles become non-convex with the emergence of double-tipped solutions for $m=1$, triple-tipped solutions for $m=2$, and so on. The double-tipped solutions were presented by Tanveer \cite{Tanveer87} for bubbles in a channel geometry while analogous families of multiple-tipped solutions were computed by Gardiner et al.~\cite{GardinerEtAlb} for the Saffman-Taylor finger problem in a channel. While our bubble solutions for $m\geq 1$ are quite possibly unstable, there is an intriguing possibility of observing multiple-tipped bubbles propagating in a Hele-Shaw cell with channel-depth perturbations using the experimental set up by Juel and collaborators \cite{deLozar2009,Thompson2014,FrancoGomez2016,Hazel2013}. From a mathematical perspective, the birth of the multiple tips as surface tension increases along a solution branch is likely to be related to exponentially small terms that appear beyond all orders of an algebraic expansion in surface tension.  We leave these issues for further research.

Finally, it would be interesting to generalise the results presented here to the case of two non-symmetric bubbles, or $p\geq3$ bubbles. For such an extension, the relevant zero surface tension solutions $z_0(\zeta)$ and $W_0(\zeta)$ are already known \cite{Crowdy2009}, and are expressed in terms of general Schottky-Klein prime functions defined over bounded $p$ connected circular domains (the intersection of the interior of the unit disc and the exterior of $p-1$ circles lying within the unit disc). The perturbation function to $z_0(\zeta)$ will take the form of a general Fourier-Laurent expansion consisting of a Taylor series and $p-1$ Laurent series about the $p-1$ interior circles. Despite the availability of fast and accurate software to compute the Schottky-Klein prime function \cite{computSK}, these more general selection problems are expected to be very computationally intensive because the loss of symmetry would imply a curvature condition akin to (\ref{steqn11}) having to be enforced on each of the $p$ boundary components. Furthermore, analogous analytical simplifications, such as those we highlighted in the case of two up-down symmetric bubbles, will not be available.



\section{Appendix}
\noindent
In this appendix, we show why (\ref{URefc}) holds true; that is
\begin{equation}
\mathrm{Re}[W_0(\zeta)+U\left(z_0(\zeta)+f(\zeta)\right)] \equiv U \mathrm{Re}[f(\zeta)], \quad \zeta \in C_0, C_1.
\label{appenid}
\end{equation}

\vskip 0.1truein
\noindent
It can be shown that function $K(\zeta;\rho)$ in (\ref{K}) satisfies the two functional identities:
\begin{equation}
K(1/\zeta;\rho)=1-K(\zeta;\rho) \quad \textrm{and} \quad K(\rho^2 \zeta;\rho)=K(\zeta;\rho)-1.
\label{kidap1}
\end{equation}
These two functional relations are all that is needed to establish (\ref{appenid}).
\vskip 0.2truein
\noindent
Now, for all $\zeta \in C_0$, $\overline{\zeta}=1/\zeta$ and we calculate $\mathrm{Re}[W_0(\zeta)]$ to be
\begin{equation}
\frac{\mathrm{i}a(1-U)}{2\sqrt{\rho}} \left[-K(-\mathrm{i}\zeta/\sqrt{\rho};\rho)-K(-\mathrm{i}\zeta \sqrt{\rho};\rho)+K(\mathrm{i}/\zeta \sqrt{\rho};\rho)+K(\mathrm{i}\sqrt{\rho}/\zeta;\rho)\right].
\end{equation}
On use of the first of the relations in (\ref{kidap1}), we have thus
\begin{equation}
\mathrm{Re}[W_0(\zeta)] = W_0(\zeta), \quad \zeta \in C_0.
\end{equation}
A similar calculation produces
\begin{equation}
\mathrm{Re}[U z_0(\zeta)] = -W_0(\zeta), \quad \zeta \in C_0.
\end{equation}
On the other hand, for all $\zeta \in C_1$, $\overline{\zeta}=\rho^2/\zeta$ and $\mathrm{Re}[W_0(\zeta)]$ can be shown to be
\begin{equation}
\frac{\mathrm{i}a(1-U)}{2\sqrt{\rho}} \left[-K(-\mathrm{i}\zeta/\sqrt{\rho};\rho)-K(-\mathrm{i}\zeta \sqrt{\rho};\rho)+K(\mathrm{i}\rho^2/\zeta \sqrt{\rho};\rho)+K(\mathrm{i}\rho^2\sqrt{\rho}/\zeta;\rho)\right].
\end{equation}
Using the second of the relations in (\ref{kidap1}),
\begin{equation}
\frac{\mathrm{i}a(1-U)}{2\sqrt{\rho}} \left[-K(-\mathrm{i}\zeta/\sqrt{\rho};\rho)-K(-\mathrm{i}\zeta \sqrt{\rho};\rho)+K(\mathrm{i}/\zeta \sqrt{\rho};\rho)+K(\mathrm{i}\sqrt{\rho}/\zeta;\rho)-2\right],
\end{equation}
followed by the first of the relations in (\ref{kidap1}), yields
\begin{equation}
\mathrm{Re}[W_0(\zeta)]=-\frac{\mathrm{i}a(1-U)}{\sqrt{\rho}} \left[K(-\mathrm{i}\zeta/\sqrt{\rho};\rho)+K(-\mathrm{i}\zeta \sqrt{\rho};\rho)\right], \quad \zeta \in C_1.
\end{equation}
A similar calculation reveals that
\begin{equation}
\mathrm{Re}[U z_0(\zeta)]=\frac{\mathrm{i}a(1-U)}{\sqrt{\rho}} \left[K(-\mathrm{i}\zeta/\sqrt{\rho};\rho)+K(-\mathrm{i}\zeta \sqrt{\rho};\rho)\right], \quad \zeta \in C_1.
\end{equation}

\vskip 0.1truein
\noindent
Thus, (\ref{appenid}) has indeed been verified.

\vskip 0.25truein
\noindent
\textbf{Acknowledgements:} CCG and SWM acknowledge the support of the Australian Research Council Discovery Project DP140100933. CCG and SWM are both appreciative of the hospitality of the School of Mathematics \& Statistics at the University of Sydney where part of this work was carried out.


\end{document}